\documentclass[12pt,final]{article}
\title{Untitled}
\author{shamit kachru}

\usepackage{hyperref}

\usepackage{cite}
\usepackage{amsmath}
\usepackage{amssymb}
\usepackage{amsthm}

\usepackage{graphicx,color}

\newcommand\nc{\newcommand}
\nc\mf\mathfrak
\nc\mc\mathcal
\nc\mb\mathbb

\begin{document}

\centerline{\large{BPS jumping loci are automorphic}}
\bigskip
\bigskip
\centerline{Shamit Kachru$^{a}$, Arnav Tripathy$^b$}
\bigskip
\bigskip
\centerline{$^a$Stanford Institute for Theoretical Physics}
\centerline{Stanford University, Palo Alto, CA 94305, USA}
\centerline{Email: skachru@stanford.edu}
\medskip
\centerline{$^b$Department of Mathematics, Harvard University}
\centerline{Cambridge, MA 02138, USA}
\centerline{Email: tripathy@math.harvard.edu}
\bigskip
\begin{abstract}
We show that BPS jumping loci -- loci in the moduli space of string compactifications where the number of BPS states jumps in an upper semi-continuous manner --
naturally appear as Fourier coefficients of (vector space-valued) automorphic forms.  For the case of $T^2$ compactification, the jumping loci are governed by a modular form studied by Hirzebruch and
Zagier, while the jumping loci in K3 compactification appear in a story developed by Oda and Kudla-Millson in arithmetic geometry.  We also comment on some curious related automorphy in the physics of black hole attractors and flux vacua.

\end{abstract}

\newpage
\tableofcontents

\section{Introduction}

The properties of BPS states are of intrinsic interest in understanding the dynamics of strongly coupled supersymmetric systems and may serve as a
useful tag in eventual classification programs.  One property of such states that has heretofore been little explored is their jumping behavior as one varies
moduli.  In \cite{KT}, motivated by understanding the jumping behavior of the Hodge-elliptic genus \cite{HEG,Wendland}, the present authors began to explore the
geometry of these loci in simple examples of string compactification.  It was found that they line up well with the notion of `special cycles in Shimura varieties,'
as we review below in \S2.

\medskip
The present work aims to describe one further striking feature of the BPS jumping loci in simple string models -- they are automorphic, in a sense which we
make precise.  The automorphy of special cycles has been developed in the mathematics literature in various series of papers.
For the simplest example of string compactification on $T^2$,  the relevant moduli space of vacua takes the form
\begin{equation}
\left( SL(2,{\mathbb Z}) \backslash \mathbb{H} \right) \times \left( SL(2,{\mathbb Z}) \backslash \mathbb{H} \right),
\end{equation}
where $\mathbb{H}$ is the usual upper half-plane $SL(2, \mathbb{R}) / SO(2)$.  (Technically, the moduli space is quotiented by the further $\mathbb{Z}_2$ of T-duality swapping the two factors.)  In the above, one factor parametrizes the complex structure of the $T^2$ and the other the (complexified) K\"ahler form.  The BPS jumping loci for a class of perturbatively visible BPS states are easily found in this case (as discussed in \cite{KT}).
As we explain in \S3, work of Hirzebruch and
Zagier on divisors in Hilbert modular surfaces \cite{HZ} precisely implies that these loci are automorphic.

\medskip
The generalization to K3 compactification is also immediate.  The moduli space of K3 compactifications of type IIA string theory takes the form
\begin{equation}
O(4,20;{\mathbb Z}) \backslash O(4,20, \mathbb{R}) / (O(4) \times O(20))
\end{equation}
while the moduli spaces of interest in complex geometry (for example, of elliptic K3s) usually take a similar form with $O(2,p)$ replacing $O(4,20)$ for various values of $p$ (for example, $p = 18$).
A series of works beginning with papers of Oda \cite{Oda} and Kudla-Millson \cite{KM}, and nicely explained in \cite{Kudla,KudlaK3}, develops a story analogous to Hirzebruch-Zagier for Shimura
varieties of this form.  The BPS jumping loci for this problem -- described in \cite{KT} as the Noether-Lefschetz loci and their generalizations to stringy geometry -- can again
be characterized as coefficients of automorphic forms.  We describe this in \S4.

\medskip
In \S5, we discuss two slightly different applications of this circle of ideas.  One application is to attractor black holes.  We explain the story, based on the
classification of attractors on $K3 \times T^2$ by Moore \cite{Mooreone, Mooretwo}, associating attractor varieties to coefficients of a modular form.  Some
version of this already appeared in the essay \cite{KTtwo}.  A second application is to flux vacua.  First, we discuss a toy model of flux counts that 
exhibits automorphy, although it does not literally arise as an example of the more general constructions in this paper.  Then, we show that the ${\cal N}=2$ flux vacua in $K3 \times T^2$
-- as described in work of Tripathy-Trivedi \cite{TT} -- are naturally counted by an automorphic form of the sort we introduced in \S4.

\medskip
We remark now that this automorphy of the jumping loci themselves is likely surprising to connoisseurs of the field. Typically, automorphic forms in this context arise as BPS partition functions; we are making a quite different claim here that the loci themselves where the BPS states exhibit certain behaviors comprise automorphic forms. It is reasonable to ask if there is any relation or common generalization of the two phenomena. We address this question in \S6 for the relatively simple case of the self-mirror type II $T^2$ compactification before making a few speculative claims for the general case.

\medskip
We close with speculations about generalizations to less symmetric vacua in \S7.

\section{Special cycles in locally symmetric spaces}

\subsection{Basic definitions}

Here, we discuss those aspects of the theory of special cycles in arithmetic locally symmetric spaces that will arise in physics
applications.  The more general story is well explained in e.g. \cite{Kudla}.

\medskip
For our purposes, an arithmetic locally symmetric space will be a double-coset space $G(\mb{Z})\backslash G(\mb{R}) / K$, where $G$ is some (reductive) group and $K$ is a maximal compact inside $G(\mb{R})$.  The theory of special cycles in such spaces has had profound importance in number theory, especially in the case when this double-coset has a natural algebraic structure (when it is said to be a Shimura variety).  For our purposes, we will immediately restrict to the case of $G$ an indefinite orthogonal group.

\medskip
Consider then a double-coset space of the form
$${\cal M}(p,q) = O(p,q;{\mathbb Z}) \backslash O(p,q, \mb{R}) / (O(p) \times O(q))~.$$
(For $p=2$, such spaces are Shimura varieties.)

\medskip
Such spaces arise commonly in parametrizing string compactifications with extended supersymmetry:

\medskip
\noindent
$\bullet$ The Narain moduli space of $T^d$ compactifications of type II strings is given by $p=q=d$.

\medskip
\noindent
$\bullet$ The Narain moduli space of heterotic $T^d$ compactifications is given by $p=d, q=d+16$.

\medskip
\noindent
$\bullet$ The moduli space of elliptic K3 compactifications of F-theory is given by the case $p=2, q=18$.

\medskip
\noindent
$\bullet$ The moduli space of IIA compactifications on K3 is given by the case $p=4, q=20$.

\medskip
\noindent
$\bullet$ The moduli space of IIB compactifications on K3 is given by $p=5, q=21$.

\medskip
\noindent
$\bullet$ The moduli space of $AdS_3 \times S^3 \times K3$ theories is given by $p=4, q=21$.

\medskip
\noindent
$\bullet$ And even some 4d ${\cal N}=2$ models have Shimura varieties arising in their moduli spaces -- for instance
$p=2,q=10$ for the vector-multiplet moduli space of the FHSV model.

\medskip
A useful way to think about these spaces, familiar from studies of worldsheet string theory, is to consider them as
moduli spaces of lattices $\Gamma^{p,q}$ of signature $(p, q)$.  Then ${\cal M}(p, q)$ parametrizes the way one
can choose ``left-moving'' and ``right-moving'' momentum sublattices of the given lattice -- in our convention, the right-movers will live in p dimensions, and left-movers in q.\footnote{We note that the
cases that arise in the simplest string theory examples are all even unimodular.}

\medskip
Note also that many more spaces may fit into the above paradigm: for example, while the Narain moduli space of toroidal heterotic compactifications is as stated, the non-perturbative moduli space for, say, $d = 7$ is also of the above form, for $p = 8, q = 24$.

\medskip
In toroidal type II compactifications, the moduli space may be corrected slightly by quotienting by a larger, extended U-duality group. For example, type II compactified on $T^6$ has the extended U-duality arithmetic group $E_{7, 7}(\mb{Z})$.  All our considerations will hold true for the double-coset space we consider, which is a cover of the true moduli space; it may well be that particularly nice automorphic forms emerge if the BPS jumping loci are grouped together more by the extended U-duality.  We leave this as an interesting question for further exploration. 

\medskip
Returning to our general double-coset space $\mc{M}(p, q)$, we now choose a vector $x$ of norm $-N$ in $\Gamma^{p,q}$, i.e. $x$ satisfying
$$\langle x, x \rangle = -N~.$$
Define a locus $D_x$ in ${\cal M}(p,q)$ as follows:
$$D_x \equiv \{ \text{locus in }{\cal M}(p,q)\text{ where }x\text{ is purely left-moving} \}.$$
${\cal M}(p,q)$ has dimension $pq$.  
The definition of $D_x$ involves precisely $p$ conditions (that $p$ dot products vanish), and so defines a sublocus in ${\cal M}$ of dimension $p(q-1)$.  In fact, this locus is abstractly a space of the form
$$O(p,q-1;{\mathbb Z}) \backslash O(p,q-1) / (O(p) \times O(q-1))~,$$
i.e. a copy of ${\cal M}(p,q-1)$ sitting in ${\cal M}(p,q)$.
This is known as a special cycle in ${\cal M}(p,q)$.
As discussed in \cite{KT}, in many concrete examples (such as those listed above) these are loci where a BPS
counting function jumps.  

\medskip
To foreshadow the sequel, we find it useful to now define a formal locus in the moduli space -- a divisor in the case of Shimura varieties -- by summing over
all such loci for a fixed $N$.  Because of the quotient by the arithmetic subgroup $O(p,q;{\mathbb Z})$, there are
a finite number of distinct vectors of norm $-N$ for each $N$, so this sum is well-defined:
$$D_N \equiv \sum_{x, \langle x,x\rangle = -N} D_x~.$$
These $D_N$ for all $N$ will play an important role in defining automorphic forms associated to BPS jumping loci.

\medskip
One can also define special cycles of higher codimension.  Here, instead of choosing a single vector $x$, one
chooses multiple vectors $x_1, \cdots, x_k$ and considers the sublocus in moduli space where the lattice generated
by these vectors becomes purely left-moving.  Clearly, for a given $k$, the relevant special cycles will now be of
dimension $p(q-k)$ and give sub varieties ${\cal M}(p,q-k) \subset {\cal M}(p,q)$.

\subsection{General philosophy of Kudla-Millson}

A fascinating feature of these special cycles, which we interpret as BPS jumping loci, is that they are automorphic. We now describe exactly what this means. For simplicity, we frequently make statements in the particular case of Shimura varieties, or $p = 2$. 
Consider first the BPS jumping locus of lowest codimension.  One can define a formal sum
$$\phi(\tau) = \sum_N [D_N] q^N~.$$
This can be viewed as defining a cohomology class in $H^2({\cal M}(2,q))$.  
For the $q^0$ term, one takes as the coefficient the first Chern class of the tautological line bundle $\omega = c_1({\cal L})$.
The striking result of Kudla-Millson is that
in fact $\phi(\tau)$ is an automorphic form; for moduli of even unimodular lattices, in fact, it is a modular form
of weight $(2+q)/2$ for $SL(2,{\mathbb Z})$.  In fact, for lattices that fail to be even unimodular, the same result holds for a congruence subgroup; for more general values of $p$, one obtains a form of weight$(p+q)/2$.  We stress again that here, the coefficients of the modular form are formal subloci, which we consider as valued in a vector space, so we have a vector-space valued modular form!  For our purposes, we simply take this vector space to be the cohomology in the appropriate codimension, but in the Shimura variety case of $p = 2$, we could refine to a Chow group-valued or even arithmetic Chow group-valued modular form.  Pairing with any functional on this vector space would return an ``ordinary'' modular form. 

\medskip
This result generalizes to BPS jumping loci of higher codimension as follows.  The data specifying a sub-lattice with
basis vectors $x_1,\cdots, x_r$, up to equivalence under the action of the arithmetic group $O(p,q;{\mathbb Z})$ 
includes the norms of the vectors and their mutual inner products.  This is $r + {r \choose 2}$ pieces of data,
naturally arranged in a symmetric $r  \times r$ matrix $M$.
One can then label a generating function for BPS jumping loci of codimension $rp$, with the data specifying a Siegel
modular form of genus $r$ for the group $Sp(r,{\mathbb Z})$.  
Again, the coefficients of the modular form are given by sums of special cycles of codimension $rp$, which share the
same data.  

\medskip
To present a more explicit formula, let us again specialize to $p=2$.  Then, the formula relevant for special cycles of higher codimension
takes the form \cite{KudlaK3}
$$\phi_r(\tau) = [\omega]^r + \sum_{rank M < r} [D^{naive}(M)] \cup [\omega]^{\rm r - rank(M)} q^M
+ \sum_{rank M = r} D[M] q^M~.$$
Here, the second term on the right hand side includes a suitable power of $\omega$ to produce a 
form in $H^{2r}({\cal M}(2,q))$.  The result of Kudla-Millson is that this object is a degree $r$ Siegel modular
form of weight ${(2+q)\over 2}$.

\section{$T^2$ compactification}

Let us proceed with a simple, explicit example.  Consider strings on $T^2$.  The torus is specified by a choice
of complex structure and complexified K\"ahler class.   In terms of the metric and B-field, one has explicitly
$$\tau = \tau_1 + i \tau_2 = {G_{12} \over G_{22}} + i {\sqrt{G} \over G_{22}}$$
$$\rho = \rho_1 + i \rho_2 = B_{12} + i \sqrt{G}~.$$

\medskip
The left and right moving momenta can be specified by choosing momenta and windings $n_{1,2}$ and $m_{1,2}$
$$p_L^2 = {1\over 2\rho_2 \tau_2} |(n_1 - \tau n_2) - \rho(m_2 + \tau m_1)|^2$$
$$p_R^2 = {1\over 2\rho_2\tau_2}|(n_1 - \tau n_2) - \bar\rho(m_2 + \tau m_1)|^2~.$$

\medskip
The duality group consists of the two $SL(2,{\mathbb Z})$ symmetries acting on the complex and K\"ahler moduli spaces together with the ${\mathbb Z}_2$ symmetries:
$$(\tau,\rho) \to (\rho,\tau)$$
$$(\tau,\rho) \to (-\bar\tau,-\bar\rho)$$
$$(\tau,\rho) \to (\tau,-\bar\rho)~.$$

\medskip
From the explicit formulae for $p_L, p_R$, we see that given a choice of integer momentum and winding quantum numbers, one can achieve $p_L=0$ for a non-trivial vector at loci in moduli space where
$$m_1 \tau\rho + m_2 \rho + n_2 \tau - n_1 = 0~.$$

\medskip
We now see a connection to the work of Hirzebruch and Zagier \cite{HZ}, where here we consider only the degenerate (split) case as opposed to an honest real quadratic field extension.  The locus $T(N)$ defined as
$$T(N) = \sum_{m_1 n_1 + m_2 n_2 = N}  \{ (z_1,z_2) \in {\cal M}(2,2); m_1 z_1 z_2 + m_2 z_2 + n_2 z_1 + n_1 = 0\}$$
is called a Hirzebruch-Zagier divisor of discriminant $N$.  This is of course a special case of the $D_N$ defined in \S2.

\medskip
The remarkable claim of \cite{HZ} concerns the formal sum
$$A(\tau) \equiv c_1(M_{-1/2}) + \sum_{m>0} T(N) q^N~.$$
Here $M_{-1/2}$ is the line bundle of modular forms of weight $-{1\over 2}$ (and
more generally $M_k$ will be the bundle of weight $k$ forms).
The theorem of Hirzebruch and Zagier states that $A(\tau)$ is a mock modular form of weight 2, valued in the
second cohomology of the moduli space $H^2({\cal M}(2,2))$.    Then in particular given any linear functional on the divisors in ${\cal M}(2,2)$, one
can naturally produce from $A(\tau)$ an ordinary mock modular form.\footnote{The theorem of Hirzebruch-Zagier has no need of mock modularity; it is only in the degenerate, split case that the severe noncompactness forces this to happen.}  We see that this is a special case of the philosophy of 
Kudla-Millson as described in \S2.2.

\medskip
Natural linear functionals include evaluating the volume form on the special cycles, and 
integrating Chern classes of $M_k$ over the cycles for various $k$.  In this particular case, 
which is a rather degenerate case of the general theory, one obtains the holomorphic part of the Eisenstein series $\hat E_2(\tau)$ (up to a prefactor) in each case
 \cite{Geer,private}:
 \begin{eqnarray*}\hat E_2(q)  &=& -{3 \over \pi \tau_2} + 1 - 24 \sum_{n=1}^{\infty} {nq^n \over 1-q^n} \\ &=& -{3\over \pi \tau_2} + 1 - 24 \sum_{n=1}^{\infty} \sigma_1(n) q^n~\end{eqnarray*}
 (where as usual $\sigma_k(n)$ denotes the sum of the $k$th powers of the divisors of $n$, $q=e^{2\pi i \tau}$, and $\tau_2$ is the 
 imaginary piece of $\tau$).
 The lack of honest modularity is due to issues of compactness.  Indeed, we need a slight extension of the Kudla-Millson philosophy in this case (and the prior case considered in the essay \cite{KTtwo}, which corresponds to the case $p = 2, q = 1$) due to the discrepancies between cohomology and compactly-supported cohomology; as usual, sufficient noncompactness violates naive modularity and instead enforces mock modularity.\footnote{We thank Jens
 Funke for substantial discussion on this issue.}

\section{K3 compactification}

Another famous case where one obtains a Shimura variety as the moduli space involves $K3$ compactification.
The moduli space of complex structures on an elliptic K3 surface is given by
$$O(2,18;{\mathbb Z}) \backslash O(2,18) / (O(2)\times O(18))~.$$
As discussed in \cite{KT}, the special cycles here -- which characterize the loci where the Picard rank jumps
from 2 (the generic value for an elliptic K3) to $2+n$ -- are sub-manifolds of the form
$${\cal M}(2,18-n) \subset {\cal M}(2,18)~.$$
 
 \medskip
 In this case, we can view the inner product on vectors as the intersection number of curves, so we characterize
 special cycles by the self-intersection of the new algebraic curve $C$ arising on the cycle
 $$C \cdot C = 2N - 2~.$$
 It follows from the general lore of Kudla and Millson that the same construction we described above --
 summing over the special cycles associated to $O(2,18;{\mathbb Z})$ inequivalent curve classes of self-intersection $2N-2$ and weighting the result by $q^N$ -- will yield Fourier coefficients of a modular form of weight 10, again valued in
 $H^{2}({\cal M}(2,18))$.
 In F-theory, this would be a counting function for loci where new BPS strings (coming from wrapped D3-branes)
 jump into existence.
 
 \medskip
 Again, we can turn this into a concrete q-series by evaluating volumes of the special cycles.  The paucity of modular
 forms of weight 10 guarantees that the result will be $\sim E_4(q) E_6(q)$:
 $$E_4(q) = 1 + 240 \sum_{n=1}^{\infty} \sigma_3(n) q^n~,$$
 $$E_6(q) = 1-504 \sum_{n=1}^{\infty} \sigma_5(n) q^n~.$$

 \section{Attractors and flux vacua}
 
 Our considerations so far have been fairly abstract, without concrete application to any particular physics question.  Here we briefly describe two physics questions which are answered by the Kudla-Millson formalism.  Our first application will be to BPS black holes in $K3 \times T^2$, and has in fact already appeared in our essay \cite{KTtwo}. We then take an interlude to discuss a toy flux model where vacua are naturally counted by automorphic forms,
 before proceeding to our second application of the formalism, which is to ${\cal N}=2$ supersymmetric flux vacua,
again in $K3 \times T^2$.
 
 \subsection{Counting attractor black holes}
 
 The attractor mechanism arises in the study of black holes in Calabi-Yau compactification of type IIB string theory \cite{FKS}.  On a 
 Calabi-Yau threefold $X$, to a charge vector
 $Q \in H^3(X,{\mathbb Z})$ and a choice of asymptotic values of vector multiplet moduli $\phi_i\vert_{\infty}$, we can associate
 an attractor point.  It characterizes the values the vector multiplet moduli flow to at the horizon of the BPS black hole
 with charge $Q$.
 
 \medskip
 For the particular case of $K3 \times T^2$ compactification, complete results about the attractor points are available from work of 
 Moore \cite{Mooreone,Mooretwo}.  The result is that each charge yields a unique attractor geometry (independent of the ``area code" or moduli at infinity).  The geometry at the black hole horizon is a combination of a `singular K3 surface'
 (one with Picard rank 20), and a CM elliptic curve.\footnote{A CM elliptic curve is one which admits complex multiplication; that is, whose complex parameter $\tau$ satisfies a quadratic equation $a\tau^2 + b\tau + c = 0$
with integer coefficients.} 
  It is further the case that by a construction of Shioda-Inose, the singular K3
 geometry can itself be encoded by an elliptic curve, which turns out to be the same CM elliptic curve!
 
 \medskip
 This gives rise to a simple application of the Kudla-Millson philosophy. The moduli space of elliptic curves is
 $${\cal M}_{2,1} = O(2,1;{\mathbb Z}) \backslash O(2,1) / (O(2) \times O(1))$$
 The CM points are the special cycles.  The discriminant of the quadratic equation that the CM curve satisfies
 plays the role of $N$ in \S2.2, and we find a mock modular form
 $$\phi_{3/2}(\tau) = \sum c_N q^N$$
 where $c_N$ counts the CM elliptic curves of discriminant $-N$.
 From the black hole perspective, $N$ controls the supergravity approximation to the black hole mass, and
 $c_N$ is counting the number of inequivalent black holes of fixed mass. 
 
 \medskip
 More detailed examination (as described in \cite{KTtwo}) shows that in this case, the form arising is in fact the mock modular form first studied
 by Zagier \cite{Zagier75},
 $$c_N = H_N$$
 with $H_N$ the Hurwitz class numbers.
 Again the mock nature of the form (where the Kudla-Millson philosophy would generally yield a holomorphic
 modular form) is due to issues of noncompactness.
 
 \medskip
 In fact, we could somehow perform essentially the same count in many different ways, each with its own physical interpretation.  For example, we could restrict to considering the complex moduli of elliptic K3s, with moduli space $\mc{M}(2, 18)$, relevant for F-theory compactification as mentioned above.  The special points should be parametrized by a Siegel modular form of degree $18$ and weight $10$, and has at least one physical interpretation as parametrizing the RCFT points in the heterotic dual frame.  Similarly, in the story above for $\mc{M}(4, 20)$, the count of attractors as special points in this moduli space should be tabulated by a Siegel modular form of degree $20$ and weight $12$.  In fact, more honestly in four dimensions, the relevant moduli space is $\mc{M}(6, 22)$, and here the count of attractors should be given by a Siegel modular form of degree $22$ and weight $14$.  Note that these forms should all essentially be Eisenstein series; this philosophy is due to the Siegel-Weil formula (the constant term of which is the famous Siegel mass formula).  Here, we may easily see that the relevant Eisenstein series, formed as usual by attempting to average over the duality group, do not converge due to insufficiently high weight; as such, all the above forms are really mock automorphic.  (As before, we may ascribe the mock nature of the form to the necessity of taking compactly-supported cohomology to obtain nontrivial counts.)
 
 \medskip
 In particular, it is natural to posit that there should be mathematical relations between all of the above forms: for example, $E_{3/2}$ should be recoverable from the Siegel-Eisenstein form of degree $20$ and weight $12$ by summing together all the coefficients with the same discriminant.  We would find it interesting to learn if there is a robust notion by which the above counts are lifts of one another.
 
\subsection{A toy model}

Next, we turn to flux compactification in IIB string theory; for a review, see e.g. \cite{DK}.
Consider string compactification on a rigid Calabi-Yau threefold $X$.  We take a symplectic basis for $H_3(X)$ to be spanned by the cycles $A$ and $B$, with the periods of the holomorphic three-form $\Omega$ are given by
$$\int_A \Omega = 1,~~\int_B \Omega = i~.$$
Our discussion of this model follows that in \S5.1 of \cite{Wati}.

\medskip
The fluxes $H_3, F_3$ can be expanded in the cohomology duals to the symplectic basis for homology, and the resulting flux superpotential takes the form
$$W = H\phi + F$$
where $\phi$ is the axio-dilaton, and
$$H = -h_1 - i h_2,~~F = f_1 + i f_2$$
with $h_{1,2}, f_{1,2} \in {\mathbb Z}$.
The tadpole for D3-charge contained in the fluxes is given by
$$N_{\rm flux} = f_1 h_2 - f_2 h_1~.$$

\medskip
The dilaton equation $D_{\phi} W =  0$ allows one to solve for
$$\bar\phi = {-{F \over H}}~.$$
Using the $SL(2,{\mathbb Z})$, we can set
$$h_1=0, 0 \leq f_2 < h_2~.$$
The D3-tadpole in the fluxes is then given by
$$N_{\rm flux} = f_1 h_2~.$$

\medskip
Let us imagine that the total tadpole we are allotted (in a suitable orientifold of $X$) is $N$; we will consider vacua of all possible $N$ abstractly,
although of course for a given threefold $X$ the number of known constructions with varying tadpole is always finite.  To satisfy tadpole cancellation,
one should then introduce $N - N_{\rm flux}$ D3-branes wandering around on $X$ and filling ${\mathbb R}^4$.  They have a moduli space of vacua
given by
$${\rm Sym}^{N-N_{\rm flux}}(\tilde X)$$
where $\tilde X$ is the moduli space of a single D3-probe of the orientifold (it is roughly given by a copy of $X$ itself, up to singularities where the brane
hits orientifold planes and so forth).

\medskip
Let us now try to estimate the number of vacua for a given $N$.

\medskip
\noindent
$\bullet$
The mobile branes would, by a standard estimate ``counting'' a given moduli space by its Euler character, yield
$$\chi({\rm Sym}^{N-N_{\rm flux}}(\tilde X))$$
vacua.

\medskip
\noindent
$\bullet$
The flux superpotential yields a unique vacuum in the axio-dilaton moduli space for each choice of $f_i, h_i$, so we can count these as 
$$\sum_{f_1 h_2 = N_{\rm flux}} h_2  = \sum_{h_2 | N_{\rm flux}} h_2 = \sigma_1(N_{\rm flux})$$
where the $h_2$ in the first expression comes from summing over possible choices of $f_2$.

\medskip
It is then natural to write the counting function for all vacua (including all possible choices of $N$), as the double sum
$$F(\sigma,\tau) = \sum_N \sum_{N_{\rm flux} \le N} p^{N_{\rm flux}} q^N \sigma_1(N_{\rm flux}) \chi({\rm Sym}^{N - N_{\rm flux}}(\tilde X))~,$$
with $p=e^{2\pi i \sigma}$ and $q = e^{2\pi i \tau}$.
This can be re-written as
$$F = \sum (pq)^{N_{\rm flux}} \sigma_1(N_{\rm flux}) q^{N-N_{\rm flux}} \chi({\rm Sym}^{N-N_{\rm flux}} (\tilde X))$$
$$\sim  {E_2(\sigma + \tau) \over \eta(\tau)^{\chi(\tilde X)}}~.$$
At the last step, we use the DMVV lift to get the eta function \cite{DMVV}, and we were sloppy with both the normalization and the
constant term in $E_2$.

\medskip
The main purpose of this intermediate section is to illustrate, in as simple a context as possible, how automorphic objects
might emerge from a microscopic attempt at counting minima of the flux potential.

\subsection{{\cal N}=2 flux vacua}
 
The conditions for ${\cal N}=2$ supersymmetric flux vacua in $X = K3 \times T^2$ compactification were described
by Tripathy and Trivedi in \S5 of \cite{TT} and studied from the perspective of gauged supergravity by 
Andrianopoli, D'Auria, Ferrara, and Lledo in \cite{Ferrara}.   The former work with the moduli space of Einstein metrics on $K3$,
$${\cal M}(3,19) = O(3,19;{\mathbb Z}) \backslash O(3,19) / (O(3) \times O(19))~,$$
while the latter work with the moduli space enlarged by further string and supergravity fields
$${\cal M}(4,20) = O(4,20;{\mathbb Z}) \backslash O(4,20) / (O(4) \times O(20))~.$$
The complex K\"ahler form $J + iB$ together with the holomorphic two-form $\Omega$ of the $K3$ and its complex
conjugate $\bar \Omega$ span a space-like three-plane, and the moduli space is the Grassmanian of
such planes in $\Gamma^{4,20}$ (i.e. in the associated real vector space).\footnote{We are mainly interested in the effects of fluxes on the K3 moduli;
the full moduli space includes additional fields that will not play a role in the sequel.}

\medskip
Type IIB flux vacua are specified by a choice of two fluxes $F_3, H_3 \in H^3(X,{\mathbb Z})$.  In this case, the
two three-forms are each given by a two-form in the $K3$ and one-form on the $T^2$.  
The results of \cite{TT,Ferrara} show that
In the special flux vacua which preserve ${\cal N}=2$ supersymmetry, the conditions
on the two-forms parametrizing the fluxes on $K3$ boil down to the choice of a vector space $V^{\rm flux}$
spanned by time-like vectors in $\Gamma^{4,20}$ and of dimension at most two, orthogonal to the
$4$-plane spanned by $\Omega, J, \cdots$.  The moduli space of such choices amounts to a copy of
${\cal M}(4,18) \subset {\cal M}(4,20).$

\medskip
These conditions are precisely those which specify a special cycle of codimension $2p$ in our general
story of \S2.2.  The special cases where $V^{\rm flux}$ is of reduced dimension are precisely the cases
which contribute to the Kudla-Millson form by multiplying powers of $\omega = c_1({\cal L})$. The three
quantum numbers characterizing a special cycle of this codimension can be given by organizing the
generators of $V^{\rm flux}$ into a $2 \times 2$ matrix.  They are the self-intersection of the RR flux,
the self-intersection of the NS flux, and the dot product of the two fluxes -- with the latter corresponding
to the ``tadpole" from the fluxes, in physics language.  Integrating the volume form on moduli space 
over the special cycles should yield a Siegel form of degree two and weight $12$.  The coefficients
of this Siegel form ``count flux vacua," in the sense that (for such a symmetric space) they are
proportional to the Euler characters of the ${\cal N}=2$ moduli spaces of flux vacua -- which is the
number of points one would expect to survive a generic perturbation lifting the moduli space.  Integrating
the volume form on these special cycles yields the Siegel-Eisenstein series of degree $2$ and weight $12$, i.e. the Eisenstein series $E_{12}(\sigma, z, \tau)$.

\section{Combining the automorphy of jumping loci and BPS counts}

We return to the $T^2$ compactification of the type II string, where as in \S3, we have \begin{eqnarray*} p_L^2 &=& \frac{1}{2 \rho_2 \tau_2} |(n_1 - \tau n_2) - \rho(m_2 + \tau m_1)|^2 \\ p_R^2 &=& \frac{1}{2 \rho_2 \tau_2} |(n_1 - \tau n_2) - \rho(m_2 + \tau m_1)|^2. \end{eqnarray*} To further simplify the moduli space, we restrict to the self-mirror locus, where we arrange conventions such that the self-mirror condition is given as $$\tau = -\overline{\rho}.$$ Note that this condition implies $\tau_2 = \rho_2$.

\medskip
Let us now consider BPS states as those for whom $p_R = 0$; for any given point $\tau$ in the moduli space $SL(2, \mb{Z}) \backslash \mb{H}$, we will take the BPS partition function (depending on $\tau$) to be given by $$BPS_{\tau}(p) = \sum_{p_R = 0} p^{p_L^2/2}.$$ We find that $p_R = 0$ if $\tau$ satisfies the equation $$m_1 \tau^2 + (m_2 - n_2) \tau + n_1 = 0.$$ For generic $\tau$, no nontrivial solutions exist and the only BPS state is the ground state, for a BPS partition function of $BPS_{\tau}(p) = 1$.  We emphasize again that here, we are counting the BPS states arising from purely left-moving vectors in the lattice; there are certainly other BPS states that exist and that we could count.  We restrict to this sector for simplicity.

\medskip
Further BPS states exist precisely at $\tau$ satisfying some quadratic equation, i.e. at the complex-multiplication elliptic curves.  Suppose $\tau$ is indeed a so-called ``quadratic irrationality'', i.e. satisfies some such quadratic equation.  This quadratic equation will be unique\footnote{The quadratic equation is only unique up to scaling, but in this analysis, we consider $D$ as fixed; if $D$ is fixed as well, then the quadratic equation is indeed fixed, up to an overall scaling by $-1$.}, so suppose we have $$a \tau^2 + b \tau + c = 0.$$ In order to have a solution in the upper half-plane, the discriminant had better be negative, so we will use the convention $$D = 4ac - b^2 > 0.$$ We then solve for $$\tau = \frac{-b + i \sqrt{D}}{2a}.$$

\medskip
Let us compute the BPS partition function for such a quadratic irrational $\tau$. As $\tau$ satisfies the unique quadratic equation $a \tau^2 + b \tau + c = 0$ but also the condition that $m_1 \tau^2 + (m_2 - n_2)\tau + n_1 = 0$, we must have\footnote{Again, up to the overall scaling by $-1$; this ambiguity only introduces a global factor of two that we here ignore.}  \begin{eqnarray*} m_1 &=& a \\ m_2 - n_2 &=& b \\ n_1 &=& c.\end{eqnarray*} We consider $m_2$ as a free variable ranging over all integers and all other momentum and winding numbers to then be uniquely fixed.

\medskip
To compute the BPS partition function, we now observe \begin{eqnarray*} p_R &=& 0 \\ \implies (n_1 - \tau n_2) &=& \overline{\rho} (m_2 + \tau m_1) \\ \implies (n_1 - \tau n_2) - \rho (m_2 + \tau m_1) &=& \overline{\rho}(m_2 + \tau m_1) - \rho(m_2 + \tau m_1) \\ &=& -2i \rho_2 (m_2 + \tau m_1) \\ \implies \frac{1}{2} p_L^2 &=& \frac{1}{4 \rho_2 \tau_2} |(n_1 - \tau n_2) - \rho(m_2 + \tau m_1)|^2 \\ &=& \frac{1}{4 \rho_2^2} |2i \rho_2 (m_2 + \tau m_1)|^2 \\ &=& |m_2 + \tau m_1|^2 \\ &=& |m_2 + \frac{-b + i\sqrt{D}}{2}|^2 \\ &=& (m_2 - b/2)^2 + D/4. \end{eqnarray*}

\medskip
Note that the parity of $b$ depends on the two cases $D \equiv 0, -1 \pmod{4}$; the first corresponds to $b$ even while the latter has $b$ odd.  This is relevant in the computation of the BPS partition function, which now depends on the quadratic irrationality $\tau$ only through the discriminant $D$.  As such, we simply write $BPS_D(p)$, which we now compute as \begin{eqnarray*} BPS_D(p) &=& \begin{cases}\hfill \sum_{m \in \mb{Z}} p^{m^2 + D/4}    \hfill & \text{ if $D \equiv 0 \pmod{4}$} \\
      \hfill \sum_{m \in \mb{Z} + 1/2} p^{m^2 + D/4} \hfill & \text{ if $D \equiv -1 \pmod{4}$} \\
  \end{cases} \\ &=& \begin{cases}\hfill \theta_3(\sigma) p^{D/4}    \hfill & \text{ if $D \equiv 0 \pmod{4}$} \\
      \hfill \theta_2(\sigma) p^{D/4} \hfill & \text{ if $D \equiv -1 \pmod{4}$.} \\
  \end{cases} \end{eqnarray*} As perhaps expected, these BPS partition functions are themselves suitably modular. 
  
\medskip
We now ask whether we can make a function which tracks both the jumping of the BPS partition function and the loci at which they jump.  To do so, we consider the quantity $$\sum_{\text{$\tau$ a quadratic irrational}} BPS_{\tau}(p) q^{D_{\tau}},$$ where $D_{\tau}$ is the discriminant associated with the quadratic irrational $\tau$. This function tracks the BPS partition function at all jump loci but also sums over all jump loci (in this case, CM points or quadratic irrationals); the $p$ fugacity tracks the extra BPS states while the $q$ fugacity tracks the jump locus.  For example, setting $p = 0$ should recover the $E_{3/2}$ count of jump loci presented in~\cite{KTtwo}.  We evaluate this function as \begin{eqnarray*} \sum_{\text{$\tau$ a quadratic irrational}} BPS_{\tau}(p) q^{D_{\tau}} &=& \sum_D BPS_D(p) H(D) q^D \\ &=& \sum_{D \equiv 0} \theta_3(\sigma) H(D) (p^{1/4} q)^D + \sum_{D \equiv 1} \theta_2(\sigma) H(D) (p^{1/4} q)^D \\ &=& \theta_3(\sigma) f_0(\sigma / 4 + \tau) + \theta_2(\sigma) f_1(\sigma / 4 + \tau),\end{eqnarray*} where here we use the notation $f_0, f_1$ for the components of $E_{3/2}$ as in (4.30) of~\cite{VW}.  Both $f_0$ and $f_1$ are mock modular forms for a subgroup of level $4$, and here we mean the holomorphic non-modular incarnation. 

\medskip
As such, this count clearly simultaneously exhibits automorphy in both variables! It is tempting to conjecture in general that such simultaneous automorphy should hold. In fact, perhaps a yet more natural object that should exhibit such simultaneous automorphy is the graded bundle of BPS states on moduli space. To be more precise, these considerations of jumping loci make it clear that the vector space of BPS states does not sweep out a bundle as one varies in moduli; instead, one expects sheaves of BPS states, perhaps one prescription for which is given by the recent proposal~\cite{Maulik}.

\section{Discussion}

In this note, we have discussed the mathematics of BPS jumping loci
in particularly symmetric string compactifications, whose moduli spaces are Shimura varieties or more generally
symmetric spaces of the form $\Gamma \backslash G / H$ with $H$ a maximal compact subgroup of $G$ and
$\Gamma$ a suitable arithmetic duality group.  We have seen that the arithmetic geometry of these loci enjoys
a beautiful theory due to Kudla-Millson, which relates the BPS jumping loci to Fourier coefficients of
automorphic forms.

\medskip
These ideas lend themselves to rampant speculation.  Precise applications to counting problems in the classes
of string vacua we mentioned here would be nice, but it is even more interesting to consider the extension to
generic ${\cal N}=2$ vacua arising from Calabi-Yau compactification.  The moduli spaces are no longer 
quotients of symmetric spaces by arithmetic groups, and the ideas used by Kudla-Millson to derive their
automorphy do not immediately generalize.  However, powerful results have been obtained in studies of 
e.g. topological string theory on the quintic Calabi-Yau by taking seriously the notion of functions automorphic
for the monodromy group defining the moduli space \cite{YY}.  It is possible that problems of characterizing
BPS jumping loci, or attractor black holes, or flux vacua in generic Calabi-Yau spaces like the quintic, will enjoy an 
analogue of the remarkable properties we discussed here.  This suggests a fascinating (if difficult) direction for
future research.

\bigskip
\centerline{\bf{Acknowledgements}}

\medskip
We thank Natalie Paquette for useful comments on a draft version of this manuscript, and we thank Jens Funke and Akshay Venkatesh for several thoroughly interesting and helpful discussions.  This work was completed
while the authors were enjoying the hospitality of the Aspen Center for Physics, supported by NSF grant
No. PHY-1066293.

\end{document}